\documentstyle[12pt,epsf]{article} \hoffset -0.5in \textwidth 6.00in \textheight
8.5in  \parskip 7pt \openup1\jot \parindent=0.5in
\newskip\humongous \humongous=0pt plus 1000pt minus 1000pt
\def\caja{\mathsurround=0pt}
\def\eqalign#1{\,\vcenter{\openup1\jot \caja
        \ialign{\strut \hfil$\displaystyle{##}$&$
        \displaystyle{{}##}$\hfil\crcr#1\crcr}}\,}
\newif\ifdtup

\def\eqright #1\cr{\noalign{\hfill$\displaystyle{{}#1}$}}
\def\eqleft #1\cr{\noalign{\noindent$\displaystyle{{}#1}$\hfill}}

\def\oldreffmt#1{\rlap{[#1]} \hbox to 2\parindent{}}

\def\figfmt#1{\rlap{Figure {#1}} \hbox to 1in{}}

\def\sectioneq{\def\theequation{\thesection.\arabic{equation}}{\let
\holdsection=\section\def\section{\setcounter{equation}{0}\holdsection}}}%

\newcounter{holdequation}

\def\auto{\eqno(\refstepcounter{equation}\theequation)}
\def\begineq #1\endeq{$$ \refstepcounter{equation}\eqalign{#1}\eqno
	(\theequation) $$}
\def\contlimit{\,{\hbox{$\longrightarrow$}\kern-1.8em\lower1ex
\hbox{${\scriptstyle (a\rightarrow0)}$}}\,}
\def\centeron#1#2{{\setbox0=\hbox{#1}\setbox1=\hbox{#2}\ifdim
\wd1>\wd0\kern.5\wd1\kern-.5\wd0\fi
\copy0\kern-.5\wd0\kern-.5\wd1\copy1\ifdim\wd0>\wd1
\kern.5\wd0\kern-.5\wd1\fi}}
\def\centerover#1#2{\centeron{#1}{\setbox0=\hbox{#1}\setbox
1=\hbox{#2}\raise\ht0\hbox{\raise\dp1\hbox{\copy1}}}}
\def\centerunder#1#2{\centeron{#1}{\setbox0=\hbox{#1}\setbox
1=\hbox{#2}\lower\dp0\hbox{\lower\ht1\hbox{\copy1}}}}
\def\lsim{\;\centeron{\raise.35ex\hbox{$<$}}{\lower.65ex\hbox
{$\sim$}}\;}
\def\gsim{\;\centeron{\raise.35ex\hbox{$>$}}{\lower.65ex\hbox
{$\sim$}}\;}

\def\super#1{\ifmmode \hbox{\textsuper{#1}}\else\textsuper{#1}\fi}
\def\textsuper#1{\newcount\holdspacefactor\holdspacefactor=\spacefactor
$^{#1}$\spacefactor=\holdspacefactor}

\def\getcite#1,{\advance\citenumber by1
\def\getcitearg{#1}\def\lastarg{@}
\ifnum\citenumber=1
\ref{#1}\let\next=\getcite\else\ifx\getcitearg\lastarg\let\next=\relax
\else ,\ref{#1}\let\next=\getcite\fi\fi\next}

\def\pom{{\rm P\kern -0.53em\llap I\,}}
\def\spom{{\rm P\kern -0.36em\llap \small I\,}}
\def\sspom{{\rm P\kern -0.33em\llap \footnotesize I\,}}

\relax

\def\auto{\eqno(\refstepcounter{equation}\theequation)}
\def\begineq #1\endeq{$$ \refstepcounter{equation}\eqalign{#1}\eqno
	(\theequation) $$}
\def\contlimit{\,{\hbox{$\longrightarrow$}\kern-1.8em\lower1ex
\hbox{${\scriptstyle (a\rightarrow0)}$}}\,}
\def\centeron#1#2{{\setbox0=\hbox{#1}\setbox1=\hbox{#2}\ifdim
\wd1>\wd0\kern.5\wd1\kern-.5\wd0\fi
\copy0\kern-.5\wd0\kern-.5\wd1\copy1\ifdim\wd0>\wd1
\kern.5\wd0\kern-.5\wd1\fi}}
\def\centerover#1#2{\centeron{#1}{\setbox0=\hbox{#1}\setbox
1=\hbox{#2}\raise\ht0\hbox{\raise\dp1\hbox{\copy1}}}}
\def\centerunder#1#2{\centeron{#1}{\setbox0=\hbox{#1}\setbox
1=\hbox{#2}\lower\dp0\hbox{\lower\ht1\hbox{\copy1}}}}
\def\lsim{\;\centeron{\raise.35ex\hbox{$<$}}{\lower.65ex\hbox
{$\sim$}}\;}
\def\gsim{\;\centeron{\raise.35ex\hbox{$>$}}{\lower.65ex\hbox
{$\sim$}}\;}

\def\super#1{\ifmmode \hbox{\textsuper{#1}}\else\textsuper{#1}\fi}
\def\textsuper#1{\newcount\holdspacefactor\holdspacefactor=\spacefactor
$^{#1}$\spacefactor=\holdspacefactor}

\def\getcite#1,{\advance\citenumber by1
\ifnum\citenumber=1
\ref{#1}\let\next=\getcite\else\ifx#1@\let\next=\relax
\else ,\ref{#1}\let\next=\getcite\fi\fi\next}

\def\upon #1/#2 {{\textstyle{#1\over #2}}}
\relax
\renewcommand{\thefootnote}{\fnsymbol{footnote}} 

\def\mainhead#1{\setcounter{equation}{0}\addtocounter{section}{1}
  \vbox{\begin{center}\large\bf #1\end{center}}\nobreak\par}
\sectioneq

\def\til#1{\centeron{\hbox{$#1$}}{\lower 2ex\hbox{$\char'176$}}}
\def\tild#1{\centeron{\hbox{$\,#1$}}{\lower 2.5ex\hbox{$\char'176$}}}
\def\sumtil{\centeron{\hbox{$\displaystyle\sum$}}{\lower
-1.5ex\hbox{$\widetilde{\phantom{xx}}$}}}

\def\pom{{\rm P\kern -0.53em\llap I\,}}
\def\spom{{\rm P\kern -0.36em\llap \small I\,}}
\def\sspom{{\rm P\kern -0.33em\llap \footnotesize I\,}}

\newcommand{\bit}{\begin{itemize}}
\newcommand{\eit}{\end{itemize}}

\newcommand{\beq}{\begin{equation}}
\newcommand{\eeq}{\end{equation}}
\newcommand{\beqa}{\begin{eqnarray}}
\newcommand{\eeqa}{\end{eqnarray}}

\begin{document} 
\begin{titlepage} 
\rightline{\vbox{\halign{&#\hfil\cr
&ANL-HEP-CP-96-74\cr 
&\today\cr}}}
\vspace{0.25in} 

\begin{center} 
 
{\Large \bf THE HARD GLUON COMPONENT OF THE QCD POMERON} 

\medskip

Alan R. White\footnote{Work supported by the U.S. Department of 
Energy, Division of High Energy Physics, Contract\newline W-31-109-ENG-38}
\\ 
\smallskip
High Energy Physics Division, Argonne National Laboratory, Argonne, IL 
60439.\\ 

\end{center}

\begin{abstract} 

We argue that deep-inelastic diffractive scaling provides fundamental
insight into the QCD Pomeron. The logarithmic scaling violations seen 
experimentally are in conflict with the scale-invariance of the BFKL 
Pomeron and with phenomenological two-gluon models. Instead the Pomeron 
appears as a single gluon at short-distances, indicating the appearance of
a Super-Critical phase of Reggeon Field Theory. That the color compensation 
takes place at a longer distance is consistent with the Pomeron carrying odd 
color charge parity. 

\end{abstract} 

\vspace{3in}
\begin{center}
Invited talk presented at the Workshop 
on Quantum Chromodynamics: Collisions, Confinement and Chaos, American 
University of Paris, France, June 3-8, 1996   
\end{center}

\renewcommand{\thefootnote}{\arabic{footnote}} \end{titlepage}

\mainhead{1. INTRODUCTION}

Deep-inelastic scaling provided the original stimulation for the development
of the parton model. Since the concept of a high-energy, short-distance,
probe of a hadron is straightforward in the deep-inelastic process, it also
has the most rigorous foundation for the application of parton model ideas
within QCD. The observed scaling violations provide much of the information
on partonic structure that is the basis for the successful application of
perturbative QCD to a wide range of hadronic physics. 

The observation of diffractive deep-inelastic scaling at HERA has opened up
a new realm of strong interaction physics. The Pomeron, which in QCD is
deeply tied to all of the long distance confinement dynamics, can now be studied
in detail at short distances. We can anticipate that diffractive scaling
violations will give crucial information on the structure of this
dynamically central part of QCD. Indeed, as we shall suggest in this talk, 
major insight into the formulation of the parton model within QCD may
actually result. The outline of the talk is as follows. 

We begin with a short summary of standard perturbative factorization, DGLAP 
evolution, and the renormalization group. Moving on to small-x, we note that
the scale-invariance of BFKL evolution implies it is not governed by the
renormalization group. Nevertheless, it can consistently appear in a
$k_{\perp}$-dependent gluon distribution and be combined with finite-order
ln[$Q^2$] scaling violations in $F_2(x,Q^2)$. 

In the diffractive cross-section $F^D_2(x,\beta,Q^2)$ there is, however, a
direct conflict between the scale-invariance property of the BFKL Pomeron
and the logarithmic scaling violations which, according to a recent H1 
analysis\cite{H1}, are present experimentally. As H1 show, their 
results require that, at short distances, the Pomeron behaves like a
single gluon (rather than the perturbative two-gluon bound state that is the
BFKL Pomeron\cite{lip}). Within QCD, gauge invariance makes this is a very
difficult property to realize. To explain how it can be achieved it is 
necessary to discuss the solution we have proposed to the full dynamical
problem of the Pomeron. 

We first recall that at long distances the (soft) Pomeron can be identified
phenomenologically as a Super-Critical Regge pole which couples to single
quarks. We describe our arguments (first put forward\cite{arw} more than fifteen
years ago) for identifying a Super-Critical phase of the Pomeron in 
QCD\cite{arw1}. In 
first approximation the Pomeron is a ``reggeized gluon'', with a dynamical
mass, in a ``reggeon condensate'' background. The condensate can be thought of 
as approximating a very soft gluon configuration accompanying the reggeized
gluon. The reggeized gluon mass scale and the condensate scale can be
distinct because they carry opposite color charge parity. The resulting
Pomeron is then distinguished from (higher-order corrections to) the BFKL
Pomeron in that it carries odd color charge parity. Necessarily, hadrons are 
not eigenstates of color parity. This is due to the appearance, in an
infinite momentum hadron, of a condensate (soft gluon) component with
non-trivial color properties. The presence of this component is also
directly related to chiral symmetry breaking and is an essential part of the
``parton'' description of an infinite momentum hadron. 

Finally we describe why, if the Pomeron is in a Super-Critical phase at 
short distances, we expect logarithmic scaling violations in deep-inelastic
diffractive scattering due to the dominance of a single hard gluon, just as
described by H1\cite{H1}. The essential feature is the separation of the
hard scale of the reggeized gluon from the soft scale of the opposite color
parity ``condensate'' (which compensates for the gluon color). 

\mainhead{2. FACTORIZATION AND DGLAP EVOLUTION}

It is well-known that in deep-inelastic scattering the operator-product 
expansion gives the leading-twist ``factorization''
$$
F_2(x,Q^2)~=~\sum_i~C_i\Bigl(x,{Q^2 \over \mu},\alpha_s(\mu^2)\Bigr)~\otimes
~f_i\Bigl(x,\mu^2,\alpha_s(\mu^2)\Bigr)~~+~~....
\auto\label{oper}
$$
where the parton densities $f_i\Bigl(x,\mu^2,\alpha_s(\mu^2)\Bigr)$ are 
matrix elements of light-cone operators. The scale $\mu$ can be used to 
separate ``hard'' from ``soft'' momenta. Application of the 
renormalization group to the $\mu$-dependence of (\ref{oper}) then allows 
the large $Q^2$ dependence of the coefficient functions $C_i$ to be determined
perturbatively via asymptotic freedom. 

The formal apparatus of the operator product expansion and the 
renormalization group is translated into simple perturbative diagrammatic 
analysis via

\begin{itemize}

\item{operator product expansion $\leftrightarrow$ factorization of collinear 
singularities}

\item{renormalization group $\leftrightarrow$ summation of logarithms via
(DGLAP) evolution equations.} 

\end{itemize}
That the solution of the DGLAP evolution equations leads to specified 
$x$-dependence, as well as $Q^2$ dependence, in deep-inelastic scattering is 
most easily seen from the ``double log'' approximation in which the 
leading ln[${1 \over x}$] contribution is kept for each log[$Q^2$]. This gives
$$
xg(x,Q^2)~~\sim~~exp\biggl[\biggl({N\alpha_s\over \pi} ln[{Q^2\over Q_0^2}]
ln[{1\over x}]\biggr)^{{1\over 2}}\biggr]
\auto\label{dlo}
$$
This approximation can be modified to comply with the 
renormalization group but to fit the small-x behavior seen in experiment, 
$Q_0$ has to be unphysically small\cite{bf}! 

\mainhead{3. BFKL EVOLUTION AT SMALL-x}

\noindent \parbox{4in} {\openup1\jot
~~~~~~~To consistently explain the small-x rise of $F_2(x,Q^2)$ within QCD 
perturbation theory we can sum the leading ln[${1\over x}$] contributions.
We sum ``reggeized gluon ladders'' via the BFKL equation\cite{lip} as 
illustrated in Fig.~1, i.e. 
$$
F_2(x,Q^2)~=~\int^1_x {dx' \over x'} \int {dQ_t^2 \over Q_t^2}
~F_2^B(x',Q_t^2,Q^2) \otimes f(x/x',Q_t^2)
\auto\label{F2x}
$$
where $f(x,k^2)$ satisfies the ``scale-invariant'' BFKL equation 
$$
{\partial \over \partial ln[{1\over x}]}f(x,k^2)~=~f^0~+~
{g^2 \over 8 \pi^3}\int {d^2k' \over {k'}^4} K(k',k)f(x,k')
\auto\label{BFKL}
$$} 
\parbox{1.95in} { \begin{center} 
\leavevmode
\epsfxsize=1.3in
\epsffile{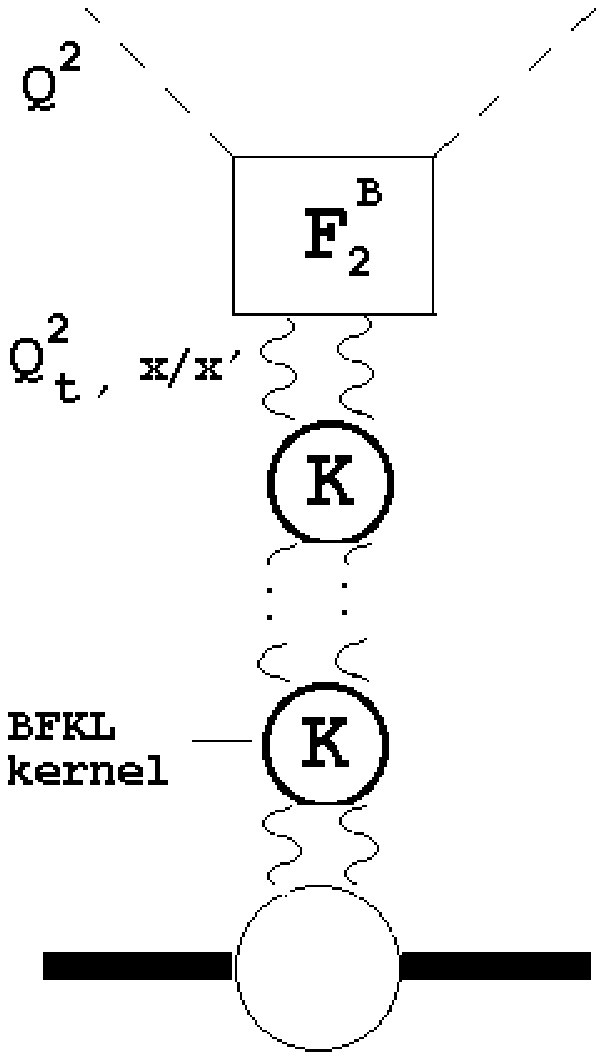}

Fig.~1
\end{center}
}
and, as illustrated, $F^B_2(x',Q_t^2,Q^2)$ comes from the quark box.

Although the scale invariance of BFKL evolution implies it can not be 
derived via the renormalization group, equations (\ref{F2x}) and 
(\ref{BFKL}) can be made consistent with DGLAP evolution by introducing 
non-perturbative splitting functions\cite{cch}. For our purposes, we can
consistently assume the BFKL small-x evolution takes place with an infra-red
transverse momentum cut-off. This (superficially at least) justifies neglect
of the evolution of the coupling and allows the cut-off to provide the scale
for the ln[$Q^2$] arising from the quark box diagram (since the cut-off will
apply also to the $Q_t$ integration in (\ref{F2x})). Phenomenologically
this implies we can describe the $Q^2$ and $x$ dependence (at small-x) by
the ``parton model'' result 
\beq\label{fit}
F_2(x,Q^2) \simeq {\alpha_s\over 3\pi}\sum_q e_q^2 x g(x)
\left({2\over 3} + \ln{Q^2\over m_g^2}\right)\ .
\eeq
where, the simple parameterization of the ``gluon distribution'' 
$g(x)=Ax^{-1-c}$ represents BFKL evolution and the ``gluon mass''
$m_g~\sim~1$ GeV is the infra-red cut-off providing the scale for the
logarithm coming from the quark box. The simple parametrization (\ref{fit})
fits the data very well. 

\mainhead{4. DEEP-INELASTIC DIFFRACTION} 

\noindent \parbox{4in} {\openup1\jot
~~~~~~~~~To study deep-inelastic diffractive scaling we ~~ consider the 
diffractive 
structure function $F^D(x_{\spom},\beta,Q^2)$ which is the large rapidity 
gap component of $F_2$. In the notation of Fig.~2 we define
$$
W^2=(P+Q)^2, ~\beta = {Q^2 \over Q^2 + M_X^2}, ~x_{\spom}={Q^2 + M_X^2
\over Q^2 + W^2}
\auto\label{var}
$$
Diffraction is a small-$x_{\spom}$ phenomenon and gluon exchanges give the
leading perturbative behavior. The} 
\noindent \parbox{1.95in} { \begin{center} 
\leavevmode
\epsfxsize=1.3in
\epsffile{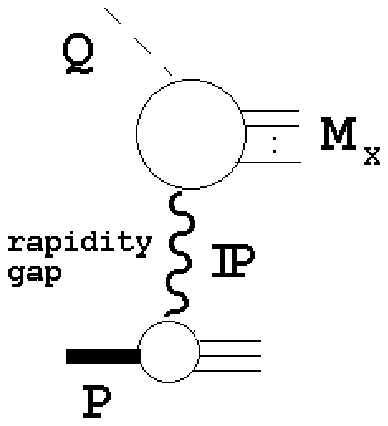}

Fig.~2
\end{center}
}
lowest-order (or BFKL Pomeron)
contribution is (color zero) two-gluon exchange coupling via a
quark-antiquark pair as in Fig.~3. 

Now the scale-invariance of the BFKL
Pomeron (or, equivalently, the infra-red finiteness due to 
gauge invariance) causes a problem. 
Consider the $k_{\perp}$ integration 
\parbox{3.5in} {\openup1\jot indicated in the diagrams of
Fig.~3. For $k^2_{\perp} \sim Q^2 >> t$ we obtain 
$$
\int {d^2k_{\perp} \over k^4_{\perp}} ~~\sim ~~{1 \over Q^2} 
\auto\label{nlt}
$$
giving non-leading twist ($1/Q^4$) behavior for the total cross-section.
Consequently the leading-twist behavior has to come from the small
$k_{\perp}$ region of the diagrams. At first sight we can treat} 
\noindent \parbox{2.45in} {\begin{center} 
\leavevmode
\epsfxsize=2.2in
\epsffile{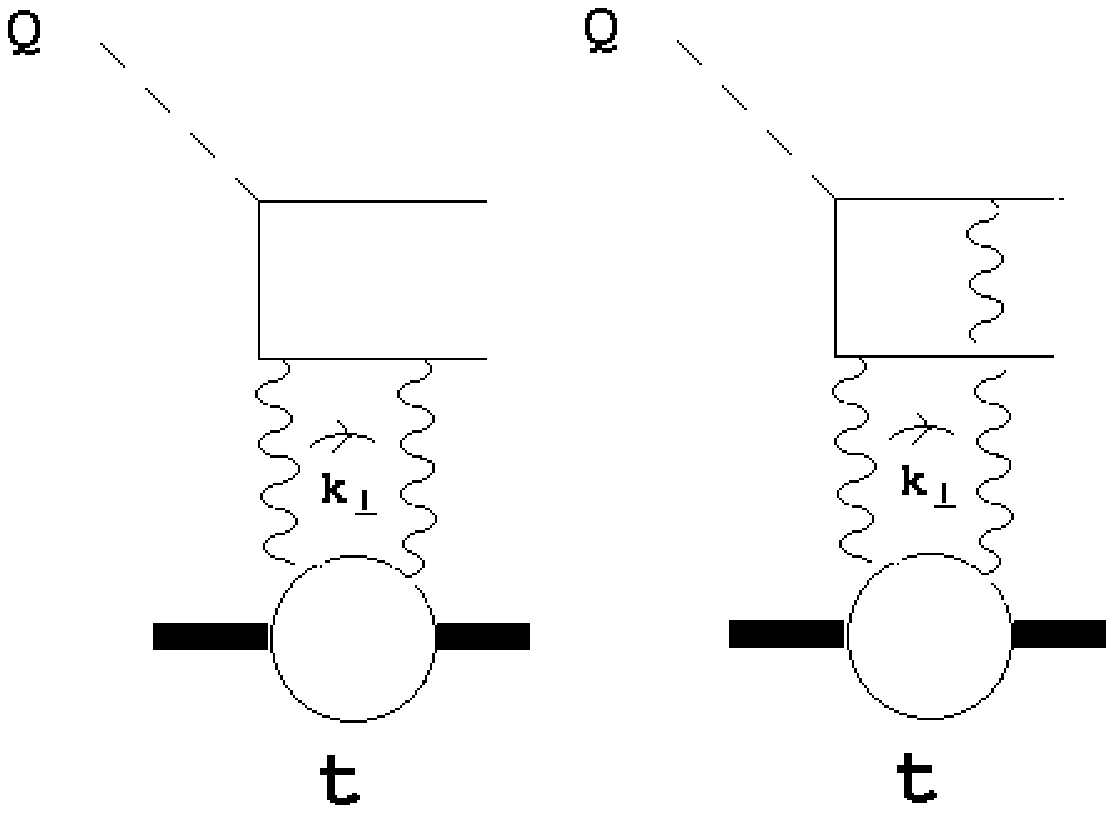}

Fig.~3
\end{center}
}
this region
with a gluon mass infra-red cut-off, as we did for the total cross-section. 
However, a manifestation of the infra-red finiteness of the BFKL Pomeron is 
that the cut-off dependence cancels between the two diagrams shown. 
Since no scale remains, the ln[$Q^2$] produced by the quark 
loop must simultaneously cancel - as indeed it 
does\cite{bw}. The same cancellation occurs in any
model which represents the Pomeron as a color zero combination of two
``non-perturbative'' gluon propagators. 

To reproduce the ln[$Q^2$] dependence seen in the H1 data\cite{H1}, it is 
necessary for one of 
the gluons in Fig.~3 to be present, with an effective infra-red cut-off, but
with the color exchange compensated by some interaction (perturbative or 
non-perturbative) associated with a smaller $k_{\perp}$ scale. This would 
allow the non-cancellation of the gluon infra-red cut-off. We would 
like to say the single ``hard'' gluon is accompanied by a cloud of
``soft'' gluons which cancels the color exchanged, but not the cut-off
dependence. However, it is gauge
invariance which produces the cut-off cancellation in the BFKL Pomeron and this
should generalise to the exchange of any number of gluons. Gauge invariance
implies all gluons in a color zero multiple gluon exchange are identical.
This argument could be avoided only if the soft gluon cloud
carries some quantum number distinguishing it from single gluon exchange -
as we describe in the next Section. 

\mainhead{5. THE SUPER-CRITICAL POMERON IN QCD}

Phenomenologically, the soft Pomeron is well-described as a Super-Critical
Regge pole, i.e. $\alpha_{\spom}(0)\equiv \alpha^0>1$. Within Reggeon 
Field Theory (RFT) it is known how this violation of unitarity is corrected 
by the summation of higher-order diagrams. Resummation gives\cite{arw1} the 
renormalized intercept $\alpha^R_{\spom}(0)<1$ together with a ``Pomeron
\parbox{4in} {\openup0.5\jot condensate''. The condensate generates new
classes of RFT diagrams whose physical interpretation is, at first, not
apparent. For example, diagrams such as that of Fig.~4 appear. The
transverse momentum poles produced by the zero energy two-Pomeron propagators 
can\cite{arw1}, in fact, be interpreted as particle poles, allowing the 
identification of the diagram as a (Pomeron) transition into two odd-signature 
reggeons, or ``reggeized gluons'' (with no color!) In this way
divergences in rapidity produced} 
\noindent \parbox{1.95in} { \begin{center} 
\leavevmode
\epsfxsize=1.7in
\epsffile{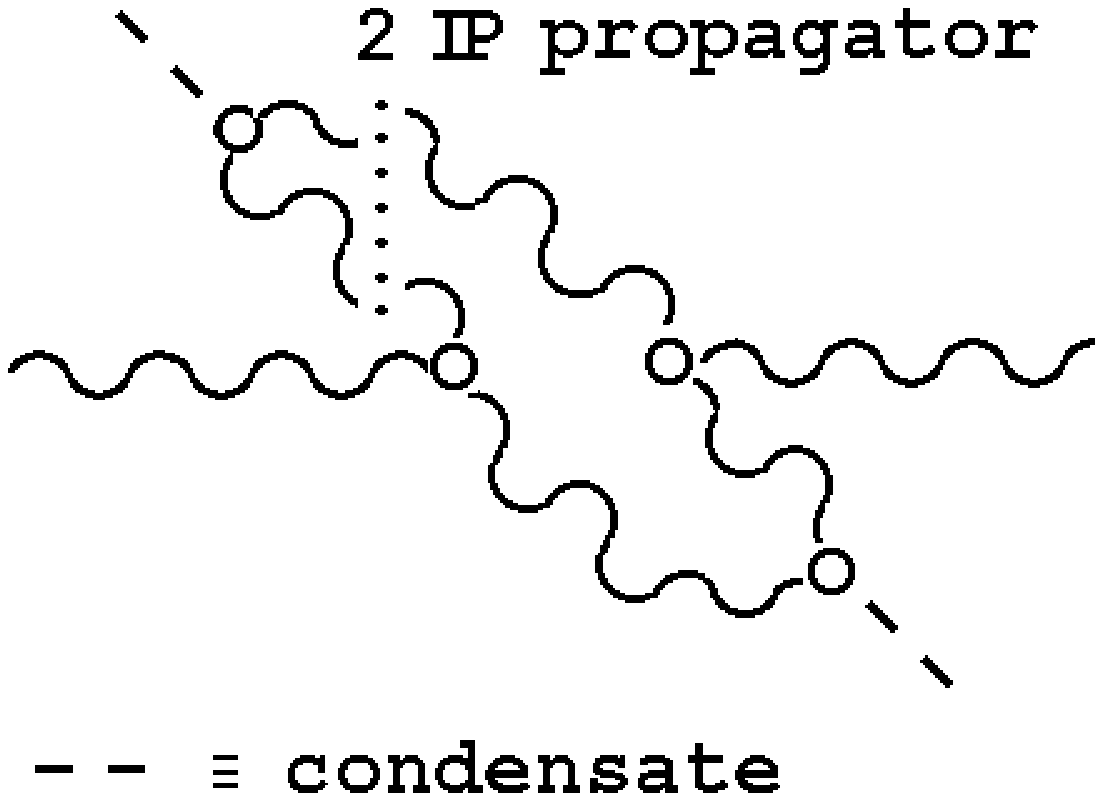}

Fig.~4
\end{center}
}
by $\alpha^0>1$ are converted to divergences in transverse momentum
associated with vector particles. Consequently the Super-Critical phase can be
characterized by the ``deconfinement of a vector gluon'' on the Pomeron
trajectory. It is very interesting to consider how this RFT phase might be
realized in QCD.

Since the RFT Critical bare intercept $\alpha^0_c>1$, it is 
possible that $\alpha^0_c>\alpha^0>1$ is the present experimental situation.
In this case the Sub-Critical expansion with $\alpha^0>1$, and no deconfined
vector gluons present, would be the right description for the 
(long-distance) soft Pomeron. Nevertheless, if we are close to
criticality, the Super-Critical phase, with a condensate and vector
``gluons'', could be present at short distances. This is how we will explain
the H1 results, as we now elaborate. 

It is not at all apparent that RFT is applicable to QCD since it requires
an isolated Regge pole as a first approximation. The BFKL Pomeron is either 
a fixed branch-point or, with $\alpha_s$ running, is an infinite set of
Regge poles. Higher-order perturbative calculations produce an even more
complicated spectrum. It is hard to understand how a spectrum of this kind
can be avoided, unless the QCD Pomeron is somehow distinguished from the
BFKL Pomeron by, for example, a quantum number of the kind that is 
needed to expose the short-distance gluon seen by H1!

In field theory, bound-state Regge poles are never isolated. Only the Regge 
poles produced by reggeization (e.g. the reggeization of the gluon) are 
isolated. Our construction\cite{arw1} of the Super-Critical Pomeron phase in
QCD builds on the reggeization of the gluon. It begins by breaking the SU(3)
gauge symmetry to SU(2) using the Higgs mechanism. In this case an SU(2) 
singlet (``deconfined'') reggeized gluon is present in the physical
spectrum. With massless quarks present there are (non-leading 
\parbox{4in} {\openup0.5\jot log) infra-red divergences present in the Regge 
limit 
of multiparticle quark/gluon scattering amplitudes which generate\cite{arw1} 
a multi-gluon condensate related to the U(1) anomaly. These divergences produce 
a confinement spectrum for the reggeon states formed. As illustrated in 
Fig.~5, the (Super-Critical) Pomeron appears as the SU(2) singlet reggeized
gluon in an ``anomalous Odderon'' condensate (i.e. a singlet  combination of
an odd number of gluons with even, rather than odd, SU(2) color} 
\noindent \parbox{1.95in} { \begin{center} 
\leavevmode
\epsfxsize=1.7in
\epsffile{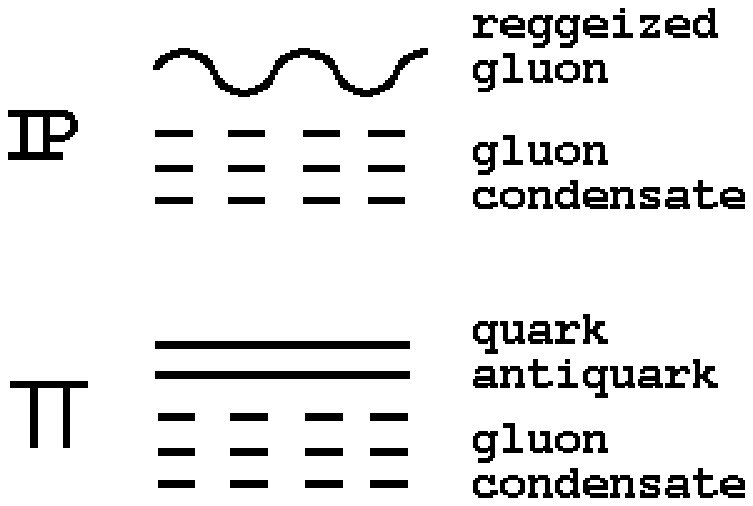}

Fig.~5
\end{center}
}
charge parity and carrying zero transverse momentum). 

The pion has a (Regge cut) odd signature constituent 
quark component, 
\parbox{4.5in} {\openup0.5\jot which the gluon
condensate converts to a pseudoscalar even signature meson, and
scattering takes place as
illustrated in Fig.~6. There is a perturbative coupling 
 of the reggeized gluon to the 
constituent quarks - as the additive quark model requires,
and the condensate self-couples (via  
massless quark loops). 
Two crucial properties of the SU(2) construction are that it
produces a Pomeron Regge pole described by RFT and that the
breaking of chiral symmetry clearly accompanies confinement.}
\parbox{1.45in} { \begin{center} 
\leavevmode
\epsfxsize=1.1in
\epsfysize=1.2in
\epsffile{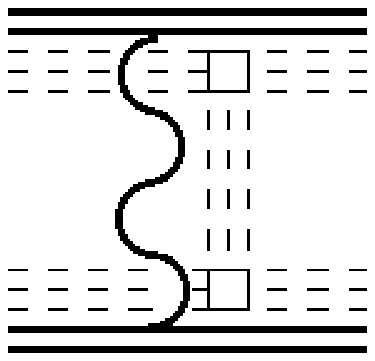}

Fig.~6
\end{center}}

To restore the gauge symmetry to SU(3) the Higgs sector must be
decoupled. (In general this requires a $k_{\perp}$ cut-off.) In this
process the condensate disappears and 
the gluons involved aquire a  $k_{\perp}$ scale. If this (cut-off dependent) 
scale becomes identical to
that of the reggeized gluon, the Pomeron effectively becomes a non-local
object in terms of gluons. However, it carries a crucial remnant of the
construction i.e. odd SU(3) color charge parity. The odd and even color 
parity, of the reggeized gluon and the condensate respectively, combine 
to give overall odd color parity. The SU(2) singlet condensate and
quark/antiquark state have projections on both SU(3) singlet and octet states 
and so, in the SU(3) limit, the pion becomes a mixture of states with even
and odd color parity (but odd physical parity). The quark-antiquark pair
appear respectively in a color singlet or a color octet combination. The odd
color parity Pomeron scatters the odd(even) state into the even(odd) state.

The construction we have outlined shows that in the Regge limit (below
some $k_{\perp}$ cut-off) QCD is describable by RFT. The BFKL Pomeron does not
contribute. 
\parbox{4.2in} {\openup0.5\jot 
(We have discussed elsewhere\cite{arw1} why the Pomeron should be
Critical if the cut-off is to be removable. We have also discussed the
relevant constraint on the quark content of the theory.) 
A-priori the full
theory may be above, or below, the critical point. As illustrated in Fig.~7,
deep-inelastic diffractive scattering will expose the simplest perturbative
contribution to the Pomeron. Odd color charge parity determines this to be
four gluon exchange in which a single antisymmetric (color) octet gluon 
combines with a
symmetric octet combination of 
three gluons. That the Pomeron }  
\parbox{1.75in} { \begin{center} 
\leavevmode
\epsfxsize=1.3in
\epsffile{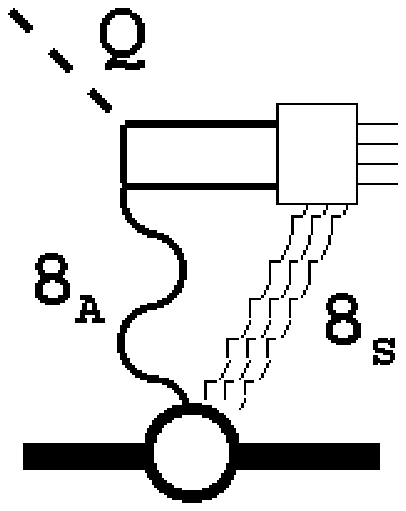}

Fig.~7
\end{center}
}
is ``in the Super-Critical
phase'' at the deep-inelastic scale implies that the symmetric octet 
forms a ``condensate'' i.e. has a lower $<k_{\perp}>$ scale than the
single gluon. As discussed in the 
last Section, this will produce the hard gluon structure seen by H1\cite{H1}.


\begin{thebibliography}{99}

\bibitem{H1} H1 Collaboration - pa02-61 ICHEP'96, Warsaw, 
Poland (1996).

\bibitem{lip} V.~S.~Fadin, E.~A.~Kuraev, L.~N.~Lipatov, {\it
Sov. Phys. JETP} {\bf 45}, 199 (1977) ; Ya.~Ya.~Balitsky and L.~N.~Lipatov,
{\it Sov. J. Nucl. Phys.} {\bf 28}, 822 (1978). 

\bibitem{arw} A.~R.~White, Proceedings of the XVIth Rencontre de Moriond, 
Vol.~2 (1981).

\bibitem{arw1} A.~R.~White, {\it Int. J. Mod. Phys.} {\bf A6}, 1859 (1991),
{\bf A8}, 4755 (1993). (The construction of the second paper is 
currently being reformulated.)

\bibitem{bf} R.~D.~Ball and S.~Forte, {\it Phys. Lett.} {\bf B335}, 77 
(1994).

\bibitem{cch} S.~Catani, M.~Ciafaloni and F.~Hautmann, {\it Nucl. Phys.} 
{\bf B366}, 135 (1991).

\bibitem{bw} J.~Bartels and M.~W\"usthoff, {\it Z. Phys.} {\bf C66}, 157 
(1995).

\end{thebibliography}
\end{document}